To be presented at *IEEE Symposium on Computational Intelligence for Financial Engineering (CIFEr)*;
Canberra, Australia, 1st-4th December 2020

# Methods Matter: A Trading Agent with No Intelligence Routinely Outperforms AI-Based Traders

Dave Cliff
*Department of Computer Science*
*The University of Bristol*
Bristol BS8 1UB, U.K.
csdtc@bristol.ac.uk

Michael Rollins
*Department of Computer Science*
*The University of Bristol*
Bristol BS8 1UB, U.K.
mr16338@bristol.ac.uk

*Abstract—* There is a long tradition of research using computational intelligence, i.e. methods from artificial intelligence (AI) and machine learning (ML), to automatically discover, implement, and fine-tune strategies for autonomous adaptive automated trading in financial markets, with a sequence of research papers on this topic published at major AI conferences such as *IJCAI* and in prestigious journals such as *Artificial Intelligence*: we show evidence here that this strand of research has taken a number of methodological mis-steps and that actually some of the reportedly best-performing public-domain AI/ML trading strategies can routinely be out-performed by extremely simple trading strategies that involve no AI or ML at all. The results that we highlight here could easily have been revealed at the time that the relevant key papers were published, more than a decade ago, but the accepted methodology at the time of those publications involved a somewhat minimal approach to experimental evaluation of trader-agents, making claims on the basis of a few thousand test-sessions of the trader-agent in a small number of market scenarios. In this paper we present results from exhaustive testing over wide ranges of parameter values, using parallel cloud-computing facilities, where we conduct millions of tests and thereby create much richer data from which firmer conclusions can be drawn. We show that the best public-domain AI/ML traders in the published literature can be routinely outperformed by a "sub-zero-intelligence" trading strategy that at face value appears to be so simple as to be financially ruinous, but which interacts with the market in such a way that in practice it is more profitable than the well-known AI/ML strategies from the research literature. That such a simple strategy can outperform established AI/ML-based strategies is a sign that perhaps the AI/ML trading strategies were good answers to the wrong question.

I. INTRODUCTION

Ever since the establishment of the first significant stock-market in Amsterdam in 1611, skilled human traders could make a lot of money by trading in financial markets, and those human traders were widely considered to be intelligent. In the past 15 years, the number of human traders at the point of execution in most of the world's major capital markets has declined sharply, as the intelligent human traders were replaced by automated *algorithmic trading* systems, colloquially known in the industry as "robot traders". Given that successful human financial-market traders were generally considered to be intelligent (for some reasonable definition of that word), there was an obvious appeal in using tools and techniques from artificial intelligence (AI) and machine learning (ML) to create smarter robot traders. Arguably the initial spark or seed for the growth of algorithmic trading was the publication of a sequence of public-domain research papers in the late 1990s and early 2000s that reported on trading algorithms which used AI/ML methods to autonomously adapt their trading behavior as market conditions altered: some of these trading algorithms were demonstrated to consistently out-perform human traders, a result that generated worldwide press coverage. Because these adaptive robot traders asked for no pay or holidays or sleep, employing profitable adaptive robot traders was clearly a more attractive commercial proposition than continuing to employ human traders, and so the reduction in human head-count began. As the amount of money to be made from automated trading grew, so did the secrecy surrounding new developments in the field: if someone creates an innovative improvement on the profitability of an automated trading system, the economically rational thing to do is to keep it a secret and quietly make money with the new technology, in the hope that none of your competitors make the same discovery.

In this paper we re-visit the sequence of papers from 1993-2008 that gave details of various public-domain trading algorithms. We describe those papers and the corresponding algorithms in more detail in Section II of this paper, but for the purposes of this introduction it suffices to say that the various algorithms are known by the acronyms ZIC, ZIP, GDX, and AA: ZIC, introduced in 1993 by Gode & Sunder [9] was an extraordinarily simple "zero intelligence" trading algorithm that could exhibit surprisingly human-like market dynamics; ZIP was publicised by Hewlett-Packard in 1997 [2] as an improvement on ZIC; GDX was released in 2001 by IBM researchers [15] as an improvement on ZIP (and was claimed at the time of its release to be the best-performing trading algorithm in the public domain); and then in 2008 Vytelingum [17] described his AA trading algorithm, which was argued to out-perform ZIP, GDX, and ZIC. In this way, a dominance-hierarchy or "pecking-order" of trading algorithms was established: if we use *A>B* to represent the claim that Algorithm *A* outperforms Algorithm *B*, then the commonly accepted view among researchers in this field, on the basis of the published research literature, was that AA>GDX>ZIP>ZIC.

In our most recent previous paper [11], we presented preliminary results which demonstrated that this commonly-accepted dominance-hierarchy among these four trading algorithms may in fact be incorrect. Our argument in [11], summarised in more detail below, was that the old results were generated from relatively simplistic experimental evaluations, and that if we apply orders of magnitude more computational resources (via the cost-reductions offered to us by contemporary cloud-computing services) and conduct truly exhaustive comparisons between these four algorithms, using a state-of-the-art market-simulator as our test-bed, we find that the dominance hierarchy in the published record is incomplete, and/or wrong. This present paper uses the same methods as we deployed in [11], but it significantly extends our prior results, and in this paper we complete the analysis started in [11]: the results in that paper, generated from hundreds of thousands of market-simulation tests, showed four pair-wise comparisons, AA-vs-ZIC, AA-vs-ZIP, GDX-vs-ZIC, and GDX-vs-ZIP, each of which was run in a pre-existing market simulator called BSE that is similar in style to the market simulators used in the original research papers that introduced each trading algorithm, but which (like the other simulators) does not accurately model the parallel and asynchronous nature of trading in real financial markets. Then the same set of experiments was repeated in a brand-new market-simulator called TBSE, which *does* incorporate parallel and asynchronous processing. In [11] we reported four sets of pairwise-comparison results from TBSE and showed that they differed in significant and interesting ways from those generated in BSE: i.e., that the nature of the market-simulator used in the comparison studies has a major effect on the dominance hierarchy. In [11] we presented tabulated quantitative data from our experiments, but in this paper we introduce graphical qualitative summaries of the numeric data, shown for our previous results in Figure 1.

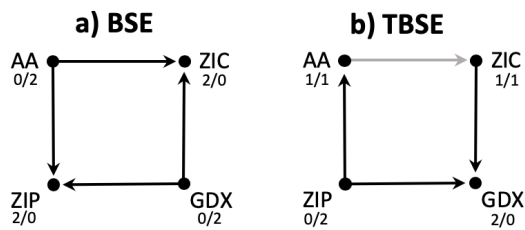

Fig. 1. Pairwise dominance graphs summarising all the results presented in our most recent paper [11]: trading algorithms AA, GDX, ZIC, and ZIP were tested in pairwise contests (i.e., A-vs-B tests) in the sequential BSE simulator (left: Fig.1a) and in the parallel and asynchronous TBSE simulator (right: Fig.1b). Nodes in the graphs represent specific trading algorithms. Each diagram shows an arrow from Algorithm A to Algorithm B if, over all sessions in the pairwise contests, A scored more "wins" than B (A "wins" a session if it scored a higher average profit per trader than B in that session). i.e. if A *dominates* B in that sense. The numbers below each strategy name are the indegree/outdegree values for that trader in this graph. In the TBSE diagram, arrows that remain the same as in BSE are shown in a pale gray, while arrows that have changed direction are shown in solid black. Note that these graphs are not fully connected, because Rollins & Cliff did not report results for AA-vs-GDX or for ZIC-vs-ZIP. Three of the four dominance relationships invert as the market-simulator test-bed is switched from the serial BSE to the parallel and asynchronous TBSE.

In this paper we present new results from extensive further experimental evaluation and comparisons, extending the four pairwise comparisons published in [11]: we show here results from AA-vs-GDX and from ZIC-vs-ZIP, which are necessary to "complete" the graphs shown in Fig.1, turning them into fully-connected networks; we then introduce two additional trading algorithms, called SHVR and GVWY, which are both minimally simple and involve absolutely no AI or ML technology: SHVR is a parasitic strategy, intended as a tongue-in-cheek model of a contemporary high-frequency trading (HFT) algorithms, which involves no intelligence other than a relentless desire to undercut all of its competitors; GVWY is, *prima facie*, a financially suicidal strategy that is hard-wired to trade at zero profit. We then run pairwise comparison experiments between GVWY and SHVR and each of AA, GDX, ZIC, and ZIP, summarising our results in a style analogous to Fig.1, but now involving fully-connected six-node networks from each of BSE and TBSE. In each of the two simulators we run one set of experiments using static market supply-and-demand curves, i.e, the kind of market situation that the original experimental evaluations of ZIC, ZIP, GDX, and AA used; and then we run another set of experiments in which the market supply and demand change dynamically throughout each market session: a situation much closer to real-world markets.

And here's the rub: counter to what many people would expect, the results we present in Section IV demonstrate that in the more realistic evaluations GVWY (which involves no computational intelligence) consistently outperforms two of the AI/ML traders, algorithms that were first described in prestigious AI publications.

We offer an explanation for why this is so, in Section V, and then in Section VI we discuss how our results highlight the extent to which a series of methodological mis-steps seems to have occurred in the sequence of influential publications that introduced these well-known trading algorithms.

Before that, Section II of this paper gives a summary of the background to this work, and is taken largely verbatim from a position paper previously published by one of us (Cliff) in 2019 [4]: our new results presented here can be read as an empirical illustration of the arguments made in that paper. Readers familiar with [4] can safely skip forward to Section III, where we describe our experiment methods in more detail. All the new experiment results presented here, and full details of the new TBSE simulator, are given in Rollins' 2020 thesis [10], which includes details of the public-domain TBSE source-code repository on GitHub: the TBSE code is freely available for inspection and use by others.

## II. BACKGROUND

### A. Traders, Markets, and Experimental Economics

The 2002 Nobel Prize in Economics was awarded to Vernon Smith, in recognition of Smith's work in establishing and thereafter growing the field of Experimental Economics (abbreviated hereafter to "EE"). Smith showed that the microeconomic behavior of human traders interacting within the rules of some specified market, known technically as an *auction mechanism*, could be studied empirically, under controlled and repeatable laboratory conditions, rather than in the noisy messy confusing circumstances of real-world markets. The minimal laboratory studies could act as useful proxies for studying real-world markets of any type, but one particular auction mechanism has received the majority of attention: the *Continuous Double Auction* (CDA), in which any buyer can announce a bid-price at any time and any seller can announce an offer-price at any time, and in which at any time any trader in the market can accept an offer or bid from a counterparty, and thereby engage in a transaction. The CDA is the basis of major financial markets worldwide, and tens of trillions of dollars flow through CDA markets every year. Understanding how best to trade in a CDA is a challenging problem. Note that here the problem of *trading* involves trying to get the best price for a transaction within the CDA.

Each trader in one of Smith's experimental CDA markets would be assigned a private valuation, a secret *limit price*: for a buyer this was the price above which he or she should not pay when purchasing an item; for a seller this was the price below which he or she should not sell an item. These limit-price assignments model the client-orders executed by sales traders in real financial markets; we'll refer to them just as *assignments* in the rest of this paper. Traders in EE experiments from Smith's onwards are often motivated by payment of some form of real-world reward that is proportional to the amount of "profit" that they accrue from their transactions: the profit is the absolute value of the difference between the limit price specified when a unit is assigned to a trader, and the actual executed transaction price for that unit: a seller profits by transacting at a price above her current assignment's limit-price; a buyer profits by transacting at a price below her current assignment's limit-price.

The limit prices in the assignments defined the market's supply and demand schedules, which are commonly illustrated in economics texts as supply and demand curves on a 2D graph with quantity on the horizontal axis and price on the vertical axis: where the two curves intersect is the market's theoretical competitive equilibrium point: indicating the equilibrium price denoted here by $P_0$. A fundamental observation from micro-economics is that competition among buyers pushes prices up, and competition among sellers pushers prices down, and these two opposing influences on prices balance out at the competitive equilibrium point; a market in which transaction prices rapidly and stably settles to the $P_0$ value is often viewed by economists as efficient (for a specific definition of efficiency) whereas a market in which transactions consistently occur at off-equilibrium prices is usually thought of as inefficient: for instance, if transaction prices are consistently above $P_0$ then it's likely that buyers are being systematically ripped off. By varying the prices in the traders' assignments in Smith's experiments, the nature of the market's supply and demand curves could be altered, and the effects of those variations on the speed and stability of the market's convergence toward an equilibrium point could be measured.

Smith's initial set of experiments were run in the late 1950's, and were described in his first paper on EE [12], published in the prestigious *Journal of Political Economy* (JPE) in 1962. The experiment methods laid out in that 1962 paper would subsequently come to dominate the methodology of researchers working to build adaptive autonomous automated trading agents by combining tools and techniques from Artificial Intelligence (AI) and Machine Learning (ML). This strand of AI/ML research converged toward a common aim: specifying an artificial agent, an autonomous adaptive trading strategy, that could automatically tune its behavior to different market environments, and that could reliably beat all other known automated trading strategies, thereby taking the crown of being the current best trading strategy known in the public domain, i.e., the *dominant strategy*. Over the past 20 years the dominant strategy crown has passed from one algorithm to another and until very recently Vytelingum's [17] *AA* strategy was widely believed to be the dominant strategy, but recent results using contemporary large-scale computational simulation techniques indicate that it does not always perform so well as was previously believed, as discussed in the next section, in which we briefly review key publications leading to the development of AA, and the recent research that called its dominance into question.

*B. A Brief History of Trading Agents*

If our story starts with Smith's 1962 JPE paper, then the next major step came 30 years later, with a surprising result published in the JPE by Gode & Sunder [9]: this popularized a minimally simple (so-called *zero-intelligence*) automated trading algorithm now commonly referred to as ZIC. ZIC traders generate their quote prices at random, using a uniform distribution that is bounded by the limit-price on their current assignment: a ZIC buyer generates quote-prices between zero and its current limit price (i.e. the limit-price is the upper-bound on the domain of the uniform distribution); a ZIC seller generates quote-prices between its limit-price (as a lower-bound) and some arbitrarily-chosen maximum allowable price.

A few years later two closely related research papers were published independently and at roughly the same time, each written without knowledge of the other: the first was a 1997 Hewlett-Packard Labs technical report by Cliff [2] describing the adaptive AI/ML trading-agent strategy known as the ZIP algorithm, inspired by ZIC but addressing situations in which markets populated by ZIC traders failed to equilibrate; the second was a 1998 paper [8] that summarized an adaptive CDA trading algorithm developed by Gjerstad as part of his economics PhD research: that trading algorithm is now widely known simply as GD. Gjerstad later worked at IBM's TJ Watson Labs where he helped set up an EE laboratory that his IBM colleagues used in a study which generated world-wide media coverage when its results were published by Das *et al.* at the prestigious *International Joint Conference on AI* (IJCAI) in 2001 [5]. This paper presented results from studies exploring the behavior of human traders interacting with GD and ZIP robot traders, and demonstrated that both GD and ZIP reliably outperformed human traders. A follow-on 2001 paper [14] by Tesauro & Das (two of the four co-authors of the IBM IJCAI paper) described a more extensively *Modified GD* (MGD) strategy, and in 2002 Tesauro & Bredin [15] then described the *GD eXtended* (GDX) strategy. Both MGD and GDX were each claimed by their IBM authors to be the strongest-known public-domain trading strategies at the times of their publication.

Subsequently, Vytelingum's 2006 PhD thesis introduced the Adaptive Aggressive (AA) strategy which again used ML techniques, and which in an *Artificial Intelligence* journal paper [17], and in later IJCAI [6] and ICAART [7] conference papers, was shown to be dominant over ZIP, GDX, and human traders. Thus far then, ZIP had been developed to improve on ZIC; ZIP had then been beaten by GDX; and AA had then beaten GDX, and hence AA held the title. In shorthand, we had AA>GDX>ZIP>ZIC.

In all of the studies discussed thus far, typically two or three different types of trading algorithm would be compared against each other on the basis of how much profit (or surplus, to use the economists' technical term) they extract from the market, so Algorithm A was said to dominate or outperform or beat or be stronger than Algorithm B if, over some number of market sessions, traders running A made more profit than traders running B. Methods of comparison varied. Sometimes a particular market set-up (i.e., a specific number of sellers, number of buyers, and their associated limit-price assignments specifying the market's supply and demand schedules) would be homogeneously populated with traders of type A, and then the same market would be re-run with all traders instead being type B, and an A-vs-B comparison of profitability in the absence of any other trading algorithms could then be made. In another design of experiment, baseline results would first be generated from a market populated homogenously by A-type traders, and then the same market experiment would be run with a single B-type trader replacing one of the A-type traders: these *one-in-many* (OIM) tests explored the attractiveness or not of traders 'mutating' or 'defecting' from using Algorithm A as their trading strategy and switching to Algorithm B. In another experiment design, for a market with D demand-side traders (buyers) and S supply-side traders (sellers), D/2 of the buyers would use Algorithm A and the remaining D/2 buyers would run Algorithm B, with the seller population being similarly split, and the A/B comparison then showed profitability in the presence of the other trading algorithm. A/B tests involving 50:50 splits, as just described, were commonly used to establish the dominance relationship between A and B. When each A-type buyer has a buy-order assignment that is also assigned to a matching B-type buyer (i.e., one A-type and one B-type buyer, each with the same limit-order assignment); and assignments are similarly "balanced" for the sellers, then the experiment design is known as a *balanced-group* (BG) test. The OIM and BG experiment designs were introduced by the IBM researchers in their work on MGD and GDX [14, 15], who proposed that BG tests are the fairest way of comparing two trading algorithms. Note that, for market with $N$ traders in it, BG tests involve a $(N/2):(N/2)$ ratio of Algorithm A to Algorithm B, while OIM tests use $(N-1):1$.

The significance of the ratio of the different trading algorithms in a test was first highlighted in Vach's 2015 master's thesis [16] which presented results from experiments with the *OpEx* market simulator [6], in which AA, GDX, and ZIP were set to compete against one another, and in which the dominance of AA was questioned: Vach's results indicate that whether AA dominates or not can in fact be dependent on the ratio of AA:GDX:ZIP in the experiment; for some ratios, Vach found AA to dominate; for other ratios, it was GDX. Vach studied only a very small sample from the space of possible ratios, but his results prompted [3] to use the public-domain *BSE* financial exchange simulator [1] to exhaustively run through a wide range of differing ratios of four trading strategies (AA, ZIC, ZIP, and the minimally simple SHVR built into BSE, explained further in Section III), doing a brute-force search for situations in which AA is outperformed by the other strategies. Cliff [3] reported on results from over 3.4 million individual simulations of market sessions, which collectively indicated that Vach's observation was correct: whether AA dominates does indeed depend on how many other AA traders are in the market, and what mix of what other strategies are also present. Depending on the ratio, AA could be outperformed by ZIP and by SHVR. Subsequent research by Snashall and Cliff [15, 16] employed the same exhaustive testing method, using a supercomputer to run more than one million market simulations (all in BSE) to exhaustively test AA against IBM's GDX strategy: this again revealed that AA does not always dominate GDX.

In this paper we will talk about counting the number of "wins" when comparing an A algorithm to a B algorithm: in the experiments reported in Section IV, we create a specific market set-up, and then run some number n of independent and identically distributed markets sessions with a specific ratio A:B of the two strategies among the buyers, and the same A:B ratio in the sellers. In any one of those sessions, if the average profit per trader (APPT) of type A traders is higher than the APPT for traders of type B, then we count that session as a "win" for A; and vice versa as a win for B.

Our preliminary experimental results reported in [12], and the much more extensive results reported here, are motivated by and extend this progression of past research. In particular, we noted that Vach's results which first revealed that the ratio of different trading algorithms could affect the dominance hierarchy came from experiments he ran using the *OpEx* market simulator [6], which is a true parallel asynchronous distributed system: OpEx involves a number of individual trader computers (discrete laptop PCs) communicating over a local-area network with a central exchange-server (a desktop PC). But many of the other results that we have just summarized came from financial-market simulators that only very roughly approximated parallel execution: the C-language source-code for the discrete-event market simulator developed to test and compare ZIC with ZIP was published in [3]; the simulators used in [14, 15] and [17] take essentially the same approach as that; and the public-domain BSE simulator [1] also uses the same very simple time-sliced approach. In all of these simulators, if any one trader is called upon to issue a response to a change in the market, it always does so within exactly one simulated time-slice within the simulation, regardless of how much computation it actually has to execute to generate that response: that is, all other activity in the market is temporarily paused, giving the currently-active trader as long as it requires to compute a response: in this sense, this style of simulator is *sequential* and *synchronous* (S&S), because even with fine time-slicing fundamentally only one trader is ever computationally active at any one time, and the computations of all other traders are paused while that one trader receives the attention of the simulator. Such S&S simulations have the net effect of treating each trader-agent's reaction-time as being so small as to be irrelevant. This approach might be defensible in human-vs-robot trading experiments, because even the slowest of these trading algorithms can compute a response substantially faster than a human can reasonably react; but if most of the traders in the market are robots, then their comparative execution times can matter a lot: if Robot A reacts faster than Robot B to some market event, then A's reaction to that market event may itself materially change the market, forcing B to re-compute its response. In this way, being faster can often be more profitable than being smarter, but an S&S simulator ignores this.

Snashall & Cliff [13] used the BSE S&S simulator to profile the reaction-times of various trading algorithms and found that they varied quite widely, indicating that a proper comparative study would require truly *parallel* and *asynchronous* (P&A) simulation test-bed, rather than an S&S one. In A P&A simulator, each robot trader is running on its own processing thread (i.e., the traders are simulated in parallel), and the computations of any one trader do not cause any other trader's computations to go into a pause-state (i.e., the traders' operations are asynchronous). In a P&A simulator, the slower traders might be expected to do much less well than when they are evaluated or compared in a temporally simplistic S&S simulation. That is what we set out to explore in the work reported here. In order to do that, one of us (Rollins) developed TBSE, a new **T**hreaded (i.e., P&A) version of the BSE financial-market simulator. TBSE is described in full in [10], and is available on the *GitHub* open-source code repository as free-to-use public-domain software. The results from our experiments, involving more than one million individual market trails, are presented in Section IV. Before that, Section III describes our methods in more detail.

III. METHODS

A. *The BSE/TBSE market-simulator test-bed*

The first version of the BSE simulator was released on *GitHub* as free-to use public-domain software in October 2012, written in *Python*. Initially developed for teaching issues in automated trading on masters-level courses at the University of Bristol, BSE has since come to be used as a reliable platform for research in automated trading: all traders in a BSE market are robot traders; the current BSE source code repository includes code for ZIC, ZIP, AA and GDX, and two additional trading algorithms called SHVR and GVWY which are described in Section III.B. BSE is written using object-oriented programming which makes it very easy for a user of BSE to add other trading algorithms, or to edit the existing ones.

As in Vernon Smith's original EE experiments, BSE allows one set of traders to be identified as buyers, and another set of traders to be identified as sellers. Any trader can be defined to run any of the trading algorithms available within BSE. Traders are issued with assignments to buy or sell via a function which can be easily switched between issuing all traders fresh assignments at the same instant (i.e., *synchronous* updating of assignments), or fresh assignments can be drip-fed into the market over the course of a market session with the assignment inter-arrival time being either regular or stochastic. Limit prices on the individual assignments (which determine the overall market's supply and demand curves, and hence its competitive equilibrium price $P_0$) can similarly be set in some regular pattern or can be randomly-generated from a specified distribution. For any particular supply and demand schedule, BSE also allows a dynamically varying *offset function* to be added to all limit-prices at the same time, where the amount added (or subtracted) from each limit-price is a function of time: in this way the relative relationship between the supply and demand curves can be maintained, while the value of $P_0$ can vary dynamically: for example, in [3], the offset function was based on time-series of real-world asset-prices, so that the BSE $P_0$ value varied over time in the same way as the original real-world asset-price. In the experiments reported here, we study the effect of varying between a null offset function (i.e. the supply and demand schedule is static for the duration of each market session, a commonplace style of experiment in all the papers reviewed in Section II.B), which we refer to as *Static $P_0$* (results given in Table I and Fig. 2); and using the sinusoidal dynamic offset function illustrated in Fig.4.8 of the BSE User Guide [1], which we refer to as *Dynamic $P_0$* (results in Table II and Fig. 3).

BSE implements a single-asset financial exchange running a CDA, and like most real-world financial exchanges it maintains a *Limit Order Book* (LOB) as the data-structure at the core of its operations. Traders in BSE issue *limit orders*, i.e. an indication of a willingness to buy or to sell: a *bid* limit order indicates an intention to buy a specified quantity of the asset at a price no greater than the price indicated on the order; an *ask* indicates a desire to sell a specified quantity at a price no lower than the price indicated on the order. Traders issue limit orders to the BSE exchange, which sorts them into the array of currently active bids, ordered best-first by price (i.e., prices sorted from highest to lowest) and the array of currently active asks, again ordered best-first by price (i.e. prices sorted from lowest to highest) the total quantity currently available at each price is also displayed, although the identities of the individual orders that collectively make up that quantity are not: in this way the exchange acts as an aggregator and anonymizer of the set of currently active limit orders. The LOB, i.e. the list of prices and quantities on the bid-side and on the ask-side, is published by the exchange to all traders in the market. At any one time the top of the LOB shows the quantity and price of the best bid, and the quantity and price of the best ask. The difference between the prices of the best bid and the best ask is known as the *spread*, and the arithmetic mean of the two best prices is the market's current *mid-price* – a common single-valued summary of current market price on a LOB.

If a trader wishes to transact at the current best price on the counterparty side of the LOB, it does so by issuing a limit order that *crosses the spread*, i.e. a bid limit order with a price higher than the current best ask (referred to as *lifting the ask*), or an ask limit order with a price lower than the current best bid (referred to as *hitting the bid*). When this happens, the quantity indicated by the spread-crossing order is consumed from the top of the LOB, and the price of the transaction is whatever best price was showing at the top of the LOB at the point that the spread-crossing order was issued. Prices in BSE are integers, representing multiples of the exchange's *tick-size*: if the tick-size is set to be $0.01 then a price of 100 in BSE represents $1.

TBSE copies BSE in all of the above regards, except where the BSE is a fine-timesliced S&S implementation of a LOB-based financial exchange, TBSE is a multi-threaded P&S implementation.

### B. Trader-agents Shaver (SHVR) and Giveaway (GVWY)

When the first version of the BSE simulator was released in 2012, the release included Python code for ZIC and ZIP, and also two additional minimally-simple built-in trading algorithms: Shaver (referred to by the ticker-style abbreviation SHVR); and Giveaway (GVWY). These were introduced initially just for illustrative purposes: SHVR is a light-hearted approximation to a high-frequency trading (HFT) algorithm, and GVWY was intended as nothing more than a stub, a bare-bones trading algorithm that other users could edit and extend. Both SHVR and GVWY are expressible in only a few, less than 10, core lines of Python code. SHVR's strategy is simple: it looks at the best price on its side of the LOB, compares it to the limit-price on its current assignment, and if it is able to then it *shaves* one tick off the best price (i.e. increasing the best bid by one penny if it is a buyer, or decreasing the best ask by one cent if it is a seller). If the shaved price would be the wrong side of the SHVR's limit price, then it does nothing. And that's it.

GVWY's strategy is even simpler. It takes the limit-price on its current assignment, and uses that as the price of its quote. It does not look at any LOB data. It only ever quotes its current assignment price. And, because the profit assigned to a trader for executing an assignment is the difference between the assignment's limit-price and the price of the transaction on that assignment, it would appear that GVWY sets out to make zero profit. But, as we will show in Section IV, and explain in Section V, in practice GVWY can be surprisingly profitable, and in our results presented here it out-performs both GDX and ZIP, casting some doubt on the actual value of those strategies.

### C. Experiment design

In this paper, as in [11], we report results from experiments in which the number of buyers $N_B$ is the same as the number of sellers $N_S$, and the supply and demand curves are randomly generated but approximately symmetric (i.e., having gradients roughly equal in magnitude but opposite in sign). We use $N_B = N_S = 20$, and run experiments at all possible A:B ratios from 1:19 through 10:10 to 19:1 – this gives us 19 different ratios to study for any one A:B pair. At any given ratio, we run 1000 independent and identically distributed market *sessions*. A single market session is a simulated version of one of Smith's CDA experiments: a clock is set to zero and starts running; traders are issued assignments according to the supply and demand functions programmed for this experiment; as time progresses traders can issue orders that are either entered on the LOB or that cross the spread and create a transaction; the BSE exchange distributes the updated LOB to each trader after any change to the LOB, and all traders react to each change in the LOB according to whatever trading algorithm they are running. Eventually the time on the simulated clock reaches the designated end-time for the session, and the session ends – at which point the average profit per trader (APPT) can be calculated for algorithm A and algorithm B, and then the "winner" of the session is the algorithm with the higher APPT.

Simulating 1000 such sessions for each of 19 ratios gives us 19,000 sessions per test of an A:B pair. Each such pair is tested in BSE and then in TBSE with static $P_0$ and then each such pair is tested again, in BSE and then in TBSE, with dynamic $P_0$, meaning that we conduct a total of 4x19,000=76,000 sessions per A/B pair, and with 15 such pairs to test this gives a total of 15x76,000=1,140,000 market sessions to be simulated.

*D. Presentation of results*

The experiment design laid out in the previous section requires us to conduct simulations of 1,140,000 individual market sessions. As we showed in [11] there is often interesting structure in the A-vs-B results when we sweep along the range of A:B ratios from one extreme to the other, but in this paper space limitations require us to skip over that structure and instead present only aggregate results, which roll up the total number of wins for A and the total number of wins for B across all the different ratios studied. As we have six trading strategies (AA, GDX, GVWY, SHVR, ZIC, and ZIP) there are 15 distinct A/B pairs to run comparative tests on. Hence for each experiment "treatment" of the equilibrium price $P_0$ (static vs dynamic) we show a 15-row table with each row showing the win-counts for A and for B in BSE and then again in TBSE. After that, we summarise the results in a network such as Fig.1.

## IV. RESULTS

Table I shows data summarising the results from running the complete set of A-vs-B tests on the six trading algorithms, in BSE and in TBSE, with a static value of the competitive equilibrium price $P_0$. Table I summarises the results from 570,000 individual market sessions. On each row of the table, the trading algorithms used as A and as B are named in the two left-most columns; then the two central columns show the number of wins for A and for B in BSE, with bold font highlighting the result from the dominant strategy (the larger of the two win-counts); and then the two rightmost columns show the win-counts for A and for B in TBSE – with bold font again highlighting the dominant result, and with underlining and italic font used to highlight those results where the dominance relationship in BSE is inverted after switching to TBSE. Fig. 2 qualitatively summarises the results of Table 1, using the graphical format established in Fig. 1.

TABLE I.    RESULTS FROM PAIRWISE CONTESTS WITH STATIC $P_0$

| AlgoA | AlgoB | BSE # A Wins | BSE # B Wins | TBSE # A Wins | TBSE # B Wins |
|---|---|---|---|---|---|
| AA | ZIC | **13824** | 5176 | **16822** | 2178 |
| AA | ZIP | **13546** | 5454 | **12441** | 6559 |
| GDX | ZIC | **13683** | 5317 | **10674** | 8326 |
| GDX | ZIP | **15684** | 3316 | *8628* | *10372* |
| ZIP | ZIC | **9930** | 9070 | **12452** | 6548 |
| AA | GDX | **11395** | 7605 | **9599** | 9401 |
| SHVR | ZIC | **13186** | 5814 | **14725** | 4275 |
| SHVR | ZIP | **14750** | 4250 | **13774** | 5226 |
| SHVR | AA | 8697 | **10303** | 8363 | **10637** |
| SHVR | GDX | **11338** | 7662 | *3927* | *15073* |
| GVWY | ZIC | **10028** | 8972 | **14198** | 4802 |
| GVWY | ZIP | **10861** | 8139 | **13005** | 5995 |
| GVWY | AA | 7402 | **11598** | 6704 | **12296** |
| GVWY | GDX | 6665 | **12335** | *11292* | *7708* |
| GVWY | SHVR | 7606 | **11394** | 7817 | **11183** |

Table II then shows data summarising the results from running the complete set of A-vs-B tests on the six trading algorithms, in BSE and in TBSE, but with the dynamically varying $P_0$ value: the format for this table is the same as that used in Table I, and again the total number of market sessions reported in Table II is 570,000, taking the total number of sessions reported in this paper to 1,140,000.

The results shown in Table II are qualitatively illustrated in Figure 3: this figure again uses the convention of pale gray being used to color arrows in the TBSE network that are unchanged from the BSE network, and it introduces an additional convention of using dashed lines to highlight dominance relationships that have changed within the relevant network as the $P_0$ treatment switches from static to dynamic.

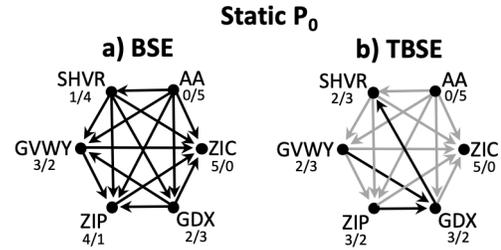

Fig. 2. Fully-connected pairwise dominance graph for the six trading strategies AA, GDX, GVWY, SHVR, ZIC, and ZIP, in BSE (left) and TBSE (right), with a static value for the competitive equilibrium price $P_0$ (i.e., quantiative data shown in Table I). Format is as for Figure 1. In BSE, it is notable that AA is a *source node*, in that it dominates all five other nodes (indegree=0); and ZIC is a *sink node* (outdegree=0), in that it is dominated by all five other nodes. In moving from BSE to TBSE, three of the dominance hierarchies reverse (as indicated by solid-black arrows). See text for further discussion.

TABLE II.    RESULTS FROM PAIRWISE CONTESTS WITH DYNAMIC $P_0$

| AlgoA | AlgoB | BSE # A Wins | BSE # B Wins | TBSE # A Wins | TBSE # B Wins |
|---|---|---|---|---|---|
| AA | ZIC | **14146** | 4854 | **16276** | 2724 |
| AA | ZIP | **15115** | 3885 | *9231* | *9769* |
| GDX | ZIC | **11039** | 7972 | *8650* | *10350* |
| GDX | ZIP | **12986** | 6014 | *5007* | *13993* |
| ZIP | ZIC | 8753 | **10247** | *17025* | *1975* |
| AA | GDX | **15420** | 3580 | *8697* | *10303* |
| SHVR | ZIC | **13419** | 5581 | **14287** | 4713 |
| SHVR | ZIP | **15875** | 3125 | *8717* | *10283* |
| SHVR | AA | 8718 | **10282** | 7221 | **11779** |
| SHVR | GDX | **15142** | 3858 | *2625* | *16375* |
| GVWY | ZIC | **10438** | 8562 | **13001** | 5999 |
| GVWY | ZIP | **11858** | 7142 | **9835** | 9165 |
| GVWY | AA | 7292 | **11708** | 8773 | **10227** |
| GVWY | GDX | **10971** | 8029 | **13538** | 5462 |
| GVWY | SHVR | 7661 | **11339** | 8430 | **10570** |

## V. DISCUSSION

The results for BSE with static $P_0$ are the closest we come here to replicating prior methods from the literature reviewed in Section II. Our results, illustrated in Fig.1a, confirm the AA>GDX>ZIP>ZIC dominance hierarchy, and the "source" nature of the AA node (zero indegree, maximal outdegree) confirms AA as the dominant algorithm, when evaluated using traditional methods. Furthermore, if the *only* zero-intelligence trader under consideration is ZIC, then the fact that the ZIC node the BSE graph on Fig.1 is a "sink" node (i.e., maximal indegree; zero outdegree) is also no surprise. But a strange feature of these BSE results is that neither GVWY or SHVR fare as badly as ZIC: GVWY dominates not only ZIC but also ZIP; and SHVR dominates *all* other strategies apart from AA. To some extent this can be explained by the fact that SHVR is essentially parasitic, relying on other traders to set a price and then merely improving on that price by a single penny: in any pairwise A-vs-B comparison involving SHVR as A and some other algorithm as B, unless B can jump ahead of SHVR in the race to a transactable price, SHVR will very often get the best of A. If you only test algorithms in a temporally simplistic (S&S) style of simulator such as BSE (which is very close to those used in the prior papers reviewed in Section II) then you might form the impression that SHVR is worth exploring further.

However, as Fig.2b makes clear, as soon as the market simulator is switched to the more realistic (P&A) style used in

TBSE, three of the dominance relationships reverse. Now the AI/ML-based traders ZIP and GDX each dominate only two other algorithms each – they both dominate ZIC, and additionally GDX now dominates SHVR, while ZIP now dominates GDX. If we rank order the algorithms on the basis of number of algorithms dominated (i.e., outdegree of their node on the graph) we have AA top-ranked with 5; SHVR and GVWY ranked second with 3 each; then ZIP and GDX with 2 each; and then ZIC with none. The fact that GDX, claimed at the time of its introduction as the best-performing public-domain trading algorithm, can be dominated in TBSE by both SHVR and GVWY is surely a sign that the original evaluations of GDX, and in particular its original market-simulator test-bed, left quite a lot to be desired.

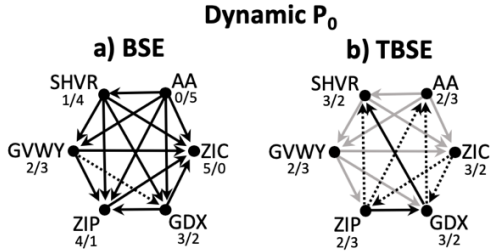

Fig. 3. Fully-connected pairwise dominance graph for the six trading strategies AA, GDX, GVWY, SHVR, ZIC, and ZIP, in BSE (left) and TBSE (right), with dynamically varying $P_0$ (i.e., quantiative data shown in Table II). Format is as for Figure 2, with the additional convention that a dotted arrow in the BSE graph in this figure is used to highlight a dominance relationship that differs in direction from the BSE graph in Figure 2; and similarly a dotted arrow in the TBSE graph here is used to highlight a difference in direction of dominance when compared to the TBSE graph in Figure 2. As in Figure 2, here the BSE graph shows AA as a source and ZIC as a sink; but in TBSE seven dominance relationships have reversed, and the graph has no pure sources or sinks.

If we now turn our attention to Fig.3, in which the market's competitive equilibrium price $P_0$ is not constant (as was often the case in the original evaluations and comparisons) but is instead dynamically varying, we see in the BSE graph of Fig.3a that the move to dynamic P0 affects only one dominance relationship (highlighted by the dashed-line arrow) when compared to the corresponding graph in Fig.2a: GDX loses its ability to dominate GVWY. However, when we examine Fig.3b we see that the switch to TBSE makes a much bigger difference: seven dominance relationships (highlighted by black-shaded arrows) invert relative to the BSE graph of Fig.3a, and four of those (highlighted by dashed-line arrows) are changed from the TBSE graph of Fig.2b. Of the four sets of results we report here, the dynamic-$P_0$ TBSE results of Fig.3b are the closest to a real-world scenario: the equilibrium price is constantly shifting, and the traders are operating in parallel and asynchronously; and, under those more realistic conditions, there are no "source" or "sink" nodes on the dominance graph and the ranking of trading algorithms by number-of-other-algorithms-dominated divides the six into two classes: the top-ranked class contains AA, GVWY, and ZIP, which each dominate three other algorithms; and the bottom-ranked class, each dominating two other algorithms, are SHVR, ZIC, and GDX, the algorithm described by IBM at the time of its introduction as the best-performing algorithmic trading system in the published literature.

The strong performance of GVWY may appear to be counterintuitive, but it is a simple consequence of how traders in any CDA-based financial exchange interact with the LOB. Because the arrival of a spread-crossing bid(ask) will lift(hit) the best ask(bid) on the counterparty side of the LOB, and the transaction then goes through at that previously-posted best price, GVWY's disregard for quoting profit-making prices means that while it will never make any profit from a limit order it quotes that does actually make it onto the LOB, very many of a GVWY's quotes will cross the spread and execute at a price *higher* than that GVWY's quoted limit-price, thereby generating a profit for the GVWY despite the fact that the GVWY's quoted price offers it no price if the transaction had gone through at that price. GVWY consistently makes money because of the way the LOB works. As it does not even involve the calculation of a randomly-chosen price, it is simpler even than ZIC, but its simplicity allows it to do well because (unlike ZIC) it will never miss the opportunity to cross the spread, if it is able to do so.

## VI. CONCLUSION

In this paper we have used the kind of comparative experimental study that is commonplace in the published literature on trading algorithms, but we have deployed resources at much greater scale than has been conventional in the past. Our results are surprising, in that they show that however interesting ZIP and GDX were at the time of their introduction, these two algorithms can be consistently outperformed by the much simpler GVWY, when the comparison takes place in a contemporary parallel and asynchronous simulator, across all possible ratios of the two trading strategies within the population of traders, and when the market session involves a dynamically varying equilibrium price. Although prior publications by other authors had presented comparisons for specific pairs (or triples) of trading algorithms, they had very often drawn conclusions from sample-sizes that were no larger than a few tens of thousands of market sessions, using supply and demand curves that gave a rarely-changing equilibrium price, and studying only one or two ratios of the traders under comparison (e.g. via one-in-many and/or balanced-group tests only). To some extent, this is probably because at the time those studies were conducted, the compute-power needed to conduct studies with sample-sizes in excess of one million would have required either a lot of money (to spend on supercomputer resources) or a lot of time (while waiting the large number of samples, of individual simulated market sessions, to complete). While lack of funds can be a tricky problem to get around, lack of time requires only patience: in principle the experiments described here could have been run 15 or more years ago; they might have locked up a few high-end PCs for some number of weeks or months, and we can only speculate that such a wait is much longer than would have been considered as acceptable at that time, by the norms of the research community. So people ran faster experiments, and reported less reliable results, results that we have now demonstrated here to be, bluntly, just wrong.

But a past lack of patience for waiting while long simulation runs conclude is only part of the problem: another manifest issue is the clear communal commitment to repeatedly re-using the experimental methods of prior researchers, traditions established long ago. If we start with the 2008 *Artificial Intelligence* paper on AA, we can see how those experiments were influenced by the 1997 paper introducing ZIP, how that ZIP paper took methodological influence from the 1993 paper introducing ZIC, and how that ZIC paper used methods commonplace in experimental economics research work stretching back all the way to Smith's 1962 *JPE* paper. Each paper used the experimental procedures and methods of one or more previous papers, which is entirely commonplace in science (we all stand on the shoulders of giants), but which led to the curious situation that the AA work published in the first decade of the 21st Century was methodologically almost a carbon-copy of Smith's seminal 1962 paper. Our work here is an attempt at nudging the field more firmly into using present-day computational methods and resources; the

Python TBSE source-code used to generate our results in this paper is now freely available on GitHub, allowing others to replicate and extend the work that we have presented here; we look forward to seeing what comes next.